\documentclass[reprint,aps,prl,amsmath,amssymb,showpacs,superscriptaddress,]{revtex4-1}
\usepackage{graphicx}
\usepackage{dcolumn}
\usepackage{bm}
\usepackage{color}
\usepackage{upgreek}

\begin{document}

\title{Mechanism of High-Order Harmonic Generation from Periodic Potentials}

\author{Tao-Yuan Du }\affiliation{State Key Laboratory of Magnetic Resonance and Atomic and Molecular Physics, Wuhan Institute of Physics and Mathematics, Chinese Academy of Sciences, Wuhan 430071, China}\affiliation{University of Chinese Academy of Sciences, Beijing 100049, China}

\author{Xue-Bin Bian}\email{xuebin.bian@wipm.ac.cn}\affiliation{State Key Laboratory of Magnetic Resonance and Atomic and Molecular Physics, Wuhan Institute of Physics and Mathematics, Chinese Academy of Sciences, Wuhan 430071, China}

\begin{abstract}

 We study numerically the Bloch electron wave-packet dynamics in periodic potentials to simulate laser-solid interactions. We introduce a quasi-classical model in the \emph{k} space combined with the energy band structure to understand the high-order harmonic generation (HHG) process occurring in a subcycle timescale. This model interprets the multiple plateau structure in HHG spectra well and the linear dependence of cutoff energies on the amplitude of vector potential of the laser fields. It also predicts the emission time of HHG, which agrees well with the results by solving the time-dependent Schr\"{o}dinger equation (TDSE). It provides a scheme to reconstruct the energy dispersion relations in Brillouin zone and to control the trajectories of HHG by varying the shape of laser pulses. This model is instructive for experimental measurements.


\end{abstract}

\maketitle

High-order harmonic generation (HHG) in atomic and molecular systems in the gas phase has been well studied theoretically and experimentally \cite{Brabec,Krausz1}. It can be understood by a semi-classical three-step model \cite{Corkum}. The bound electron may be ionized by tunneling, then driven by the external laser field. When it recombines with the parent ion, harmonics are emitted. The cutoff energy is around $I_{p}$+3.17$U_{p}$ ($U_{p}$ is the ponderomotive energy, which is proportional to $A_{0}^{2}$, where $A_{0}$ is the amplitude of the vector potential of the laser fields). The HHG has resulted in the birth of attosecond (1 as = $10^{-18}$ s) pulses \cite{Goulielmakis,Zhao} and new imaging tools, such as molecular tomography \cite{Itatani} and spectroscopy \cite{Worner}. Recent experiments have demonstrated that light-solid interactions offer a wide range of other phenomena and applications to be explored \cite{Schultze,Schiffrin,Kruger}, including HHG from solid-state materials \cite{Ghimire,Schubert,Luu,Hohenleutner,Langer}.

Experimental results present a recollision feature in two-color laser fields \cite{Vampa3}, which is similar to HHG from the gas phase. However, it is also found that the cutoff of HHG from the solid phase depends on the strength $E_{0}$ of the laser fields linearly \cite{Ghimire}, rather than quadratically in the gas phase. Theoretical studies \cite{Wu,Guan} show that the cutoff energy of HHG from the solid phase also depends on the laser wavelength $\lambda$ linearly, rather than $\lambda^{2}$ in the gas phase. Multi-plateau structure in solid HHG has also been found theoretically \cite{Wu,Guan} and experimentally \cite{Ndabashimiye} recently, which is quite different from HHG in the gas phase. Two-band \cite{Vampa4} and multiband models \cite{Wu,Guan}, intraband \cite{Ghimire2,Hawkins2} and interband transitions \cite{McDonald} are used to explain the mechanism behind HHG. However, simple quantitative universal models are still required to reveal the mechanisms behind HHG.

This work introduces a quasi-classical model to investigate the electron dynamic processes under the laser fields in the wave vector \emph{k} space, which is similar to the three-step model for HHG generated from the atomic and molecular systems in the coordinate space \cite{Corkum}.

It is also described into three steps: Zener tunneling, electron wave packet oscillation in conduction bands, and an instantaneous electron-hole pair recombination. Since the electron and hole are delocalized, it is unnecessary for them to return to their original positions to emit HHG. The interband nonresonant and resonant Zener tunneling had been studied previously \cite{Golde,Glutsch,Hawkins2,Wismer}. The electron-hole recollisions have also been observed in experiments \cite{Zaks,Langer}. Here, we make an approximation that the population of the valence band is only weakly depleted. In addition, the durations of the tunnel and recombination processes are neglected.

Due to the drive of laser fields, electrons in the valence band have probabilities to tunnel to conduction bands, i.e. Zener tunneling. But the tunneling probabilities exponentially decay with the increase of energy gap. Only a small portion of electrons populated near the wave vector $k=0$ on top of the valence band with minimal band gap can tunnel to conduction bands with the laser parameters used in the current work. So we choose an initial state with $k_0=0$.

\begin{figure}
\centering\includegraphics[width=8 cm,height=10 cm]{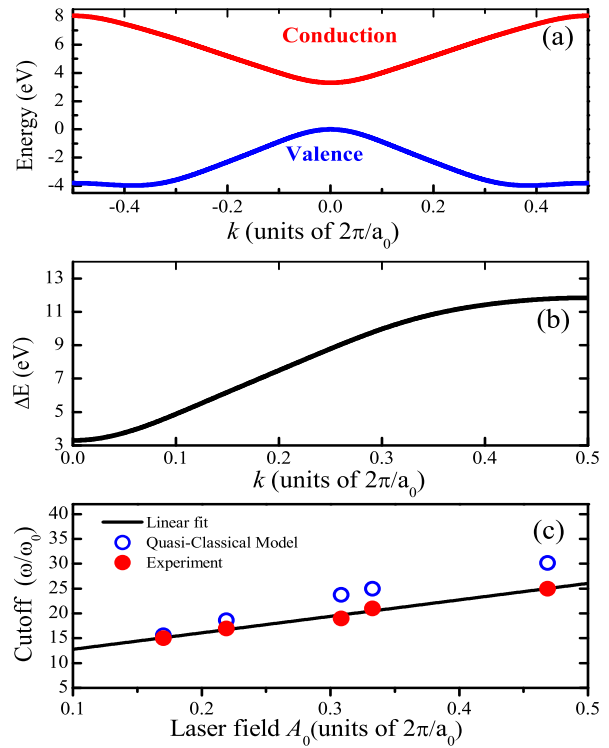}
\caption{(Color online) (a) The first conduction (red line) and the valence (blue line) bands along the $\Gamma$-M direction of the Brillouin zone from Ref. \cite{Vampa4,Goano}. (b) The band gap between conduction and valence bands along the $\Gamma$-M as a function of wave vector \emph{k}. (c) Comparison of the cutoff energies of HHG from the experimental results \cite{Ghimire} and the proposed quasi-classical model.}\label{Fig1}
\end{figure}

After tunneling into the conduction bands, the Coulomb forces are neglected, which is similar to the strong-field approximations in the gas phase. The motion of Bloch electron wavepacket in the wave vector \emph{k} space within conduction bands is given by

\begin{equation}\label{E1}
k(t) = k_0 + \frac{e}{\hbar}A(t),
\end{equation}
\emph{A}(t) is the vector potential of the laser field. Although the internal interactions in solids are neglected, the energy band structure is embedded in this model, i.e., the energy of the electron is not $\frac{k^2}{2}$, but $\epsilon(k(t))$ computed based on Bloch theorem. The HHG emissions synchronize with the Bloch electron wave packet motions. The HHG orders $\eta$ is determined by
\begin{equation}\label{E2}
\eta(t) = \frac{\epsilon_{c}(k(t))-\epsilon_{v}(k(t))}{\hbar\omega_{0}},
\end{equation}
where $\omega_{0}$ is the laser field frequency, $\epsilon_{c}$ and $\epsilon_{v}$ represent the energies of the conduction and valence bands, respectively.
Based on this quasi-classical model, the cutoff energy $\eta_{cutoff}$ is determined by the maximal band gap between the conduction (including higher bands) and valence bands in the range of $\emph{k}\in[0,k_{max}]$. For ZnO, its energy band structure has been well studied as illustrated in Fig. 1(a). One may find that the energy gap ($\triangle E$) between the first conduction band and the valence band is linearly dependent on \emph{k} approximately in the range of [0, 0.4] a.u. as shown in Fig. \ref{Fig1}(b).
Due to the fact that the Bloch electron wave-packet motion has a maximal displacement $k_{max}$ in the wave vector space, the cutoff energy depends linearly on the amplitude $A_0$ from Eq.(\ref{E1}) if we adopt the initial $k_0=0$. As a result,
\begin{equation}\label{E3}
\eta_{cutoff} \propto  \triangle\epsilon_{k} \propto k_{max} \propto A_{0} \propto E_{0}\lambda.
\end{equation}

In turn, once the parameters of driving fields and $\eta_{cutoff}$ are given, the energy bands can be reconstructed by this model.

The cutoff energies by the above quasi-classical model compared with experimental data \cite{Ghimire} are presented in Fig. \ref{Fig1}(c). The energy bands of bulk crystal ZnO are reproduced from Ref. \cite{Vampa4,Goano}. To establish the relation between electric fields of inside and outside of the solids, we need to take into account the reflection of incident pulses at the surface. The relation between the incident electric field from outside the medium, $E_{in}$, and the field in the medium, $E_{medium}$, is given by $E_{medium}=\frac{2}{1+n}E_{in}$ \cite{Vampa4}. \emph{n} represents the refractive index and is set to be 1.8992 for the laser fields with wavelength 3.25 $\mu$m adopted in the experiment. The $E_{medium}$ was used in the calculations of quasi-classical model in Fig. \ref{Fig1}(c).

The cutoff energies in this model agree well with experimental results when the intensity is low. A cutoff frequency discrepancy of 4-5 orders can be observed for a higher intensity. This may come from the bigger ionization rate and fast decay of laser intensity during the propagations in the crystal. Another possible well-known reason is that accurate laser intensity is not easy to be measured experimentally. One can draw a conclusion that the quasi-classical model based on two bands transitions is qualitatively in accord with the experimental measurements.

The experimental data about cutoff dependence on wavelength are less. Next, in order to further verify this quasi-classical model, quantum simulations involving multiple bands are performed.
We describe the light-solid interaction in one dimension, along the polarization direction of the laser fields. In the velocity-gauge treatment, the time-dependent Hamiltonian is written as
\begin{equation}\label{E4}
\hat{H}(t)=\hat{H}_0 + \frac{e}{m}A(t)\hat{p},
\end{equation}
where $\hat{H}_0=\frac{{\hat{p}^2}}{2m}+V(x)$, and $V(x)$ is a periodic lattice potential. In our calculations, we choose the Mathieu-type potential \cite{Slater}. The specific form is $V(x)=-V_0[1+\cos(2\pi x/a_0)]$, with $V_0=0.37$ a.u. and lattice constant $a_0=8$ a.u. and $\hat{p}$ is the momentum operator.

The energy band structure and time-dependent Schr\"odinger equation (TDSE) can be solved by using Bloch states in the $k$ space and $B$-spline functions in the coordinate space respectively. For details we refer readers to Refs. \cite{Wu, Guan}. After obtaining the time-dependent wave function $|\psi(t)\rangle$ at an arbitrary time, we can calculate the laser-induced currents by:

\begin{equation}\label{E5}
j(t)=-\frac{e}{m}\Big( Re\left[\langle \psi(t)|\hat{p}|\psi(t)\rangle\right] + eA(t) \Big).
\end{equation}

The HHG power spectrum is proportional to $|\emph{j}(\omega)|^{2}$, the modulus square of Fourier transform of the time-dependent current in Eq. (\ref{E5}).

\begin{figure}
\centering\includegraphics[width=8 cm,height=11 cm]{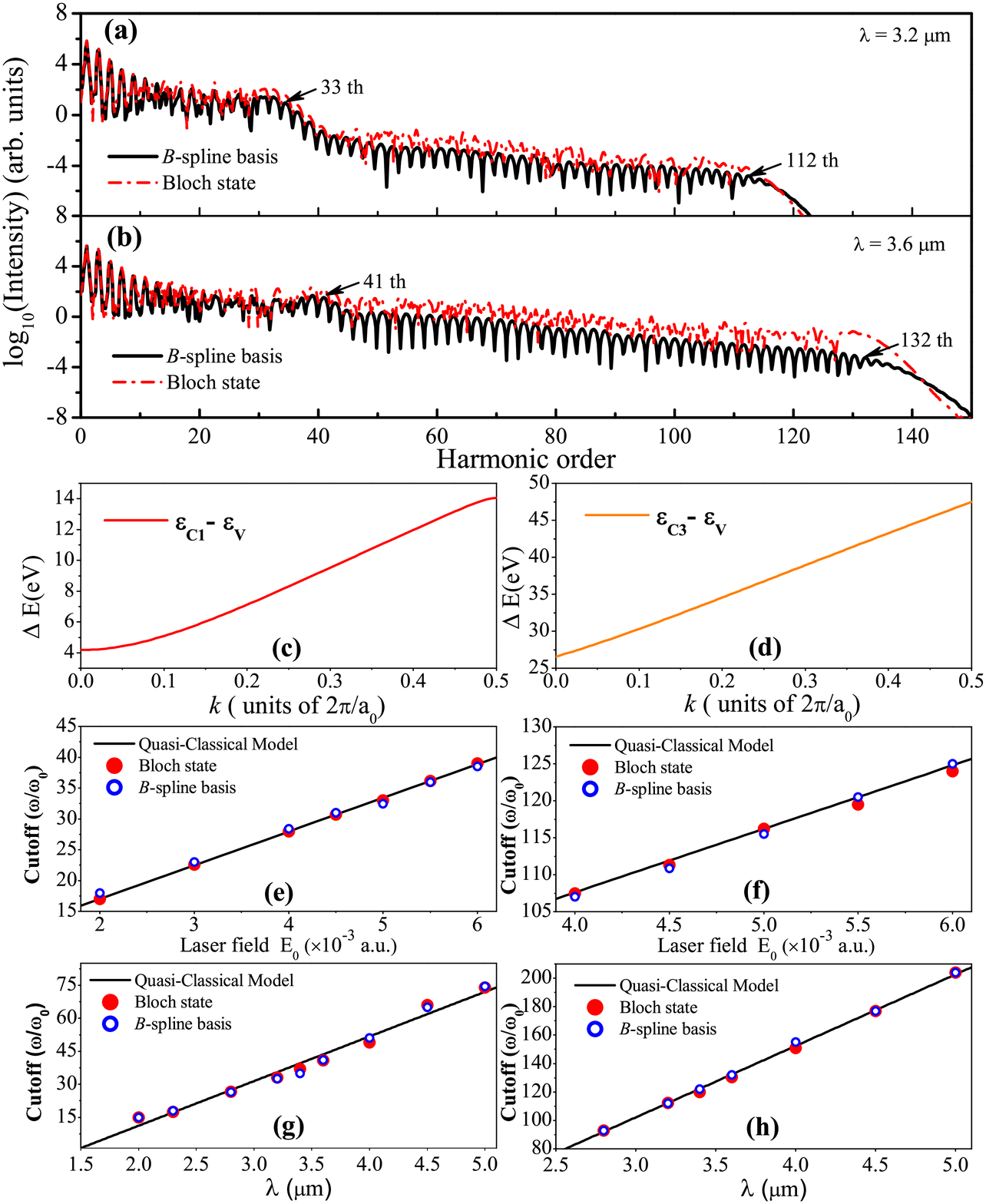}
\caption{(Color online) HHG spectra and field dependence of the cutoff. (a) and (b) show the whole HHG spectra calculated by Bloch states and \emph{B} splines, the wavelengths of laser fields are 3.2 $\mu$m and 3.6 $\mu$m, respectively. The strength of the laser fields is fixed at 0.005 a.u. The arrows show the cutoffs calculated by the quasi-classical model. (c) and (d) show the band gap between conduction band C1, C3 and valence band, respectively. (e) and (f) show the field strength dependence of cutoff for the first and second plateaus, respectively. The wavelength of laser fields is fixed at 3.2 $\mu$m. (g) and (h) show the wavelength dependence of cutoff for the first and second plateaus, respectively. The strength of laser fields is fixed at 0.005 a.u. }\label{Fig2}
\end{figure}

\begin{figure}
\centering\includegraphics[width=8 cm,height=12 cm]{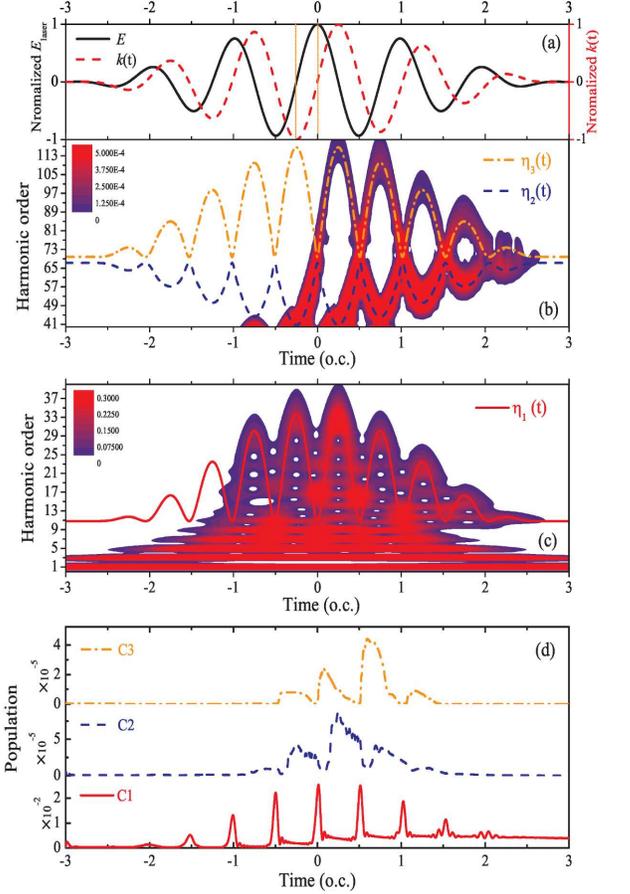}
\caption{(Color online) Time-Frequency analysis of HHG. The colormap is extracted from the quantum TDSE simulations, while the dash dot curves represent the predictions by the quasi-classical model. (a) the time-dependent displacement \emph{k}(t) in \emph{k} space and electric field \emph{E} of the laser pulses. The values have been normalized. (b) and (c) are the time-frequency analysis of the second and first plateaus, respectively.
$\eta_1$, $\eta_2$ and $\eta_3$ represent the results in quasi-classical model involving the conduction band C1, C2 and C3, respectively. (d) Populations of conduction bands C1, C2 and C3.
Laser parameters are the same as those in Fig. \ref{Fig2} (a). The envelope of the pulse is a $\cos^2$ function, and the total duration is 6 cycles.}\label{Fig3}
\end{figure}

\begin{figure}
\centering\includegraphics[width=8 cm,height=9 cm]{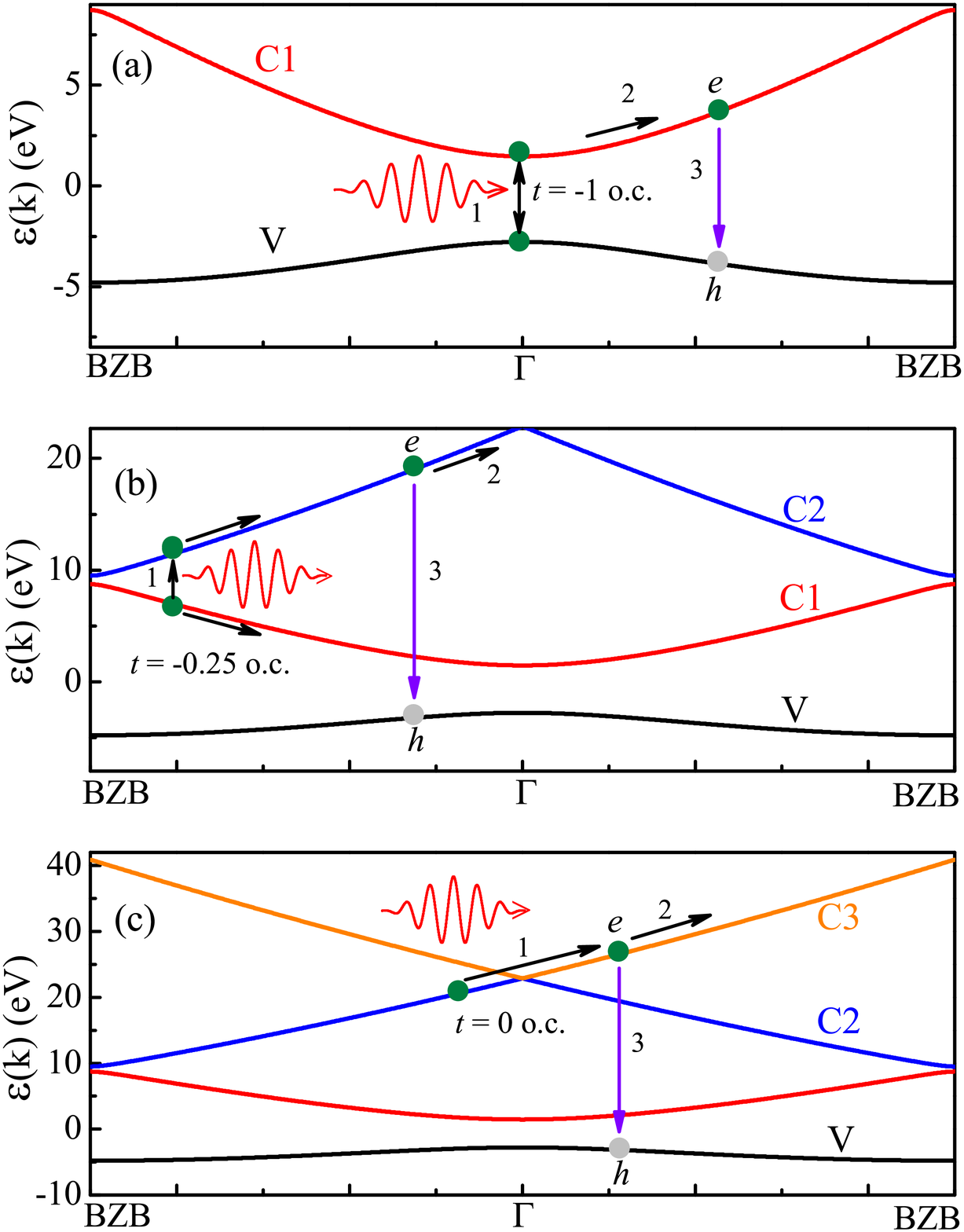}
\caption{(Color online) Microscopic motion within sub-cycle timescale and electron-hole pair recombination in energy bands driven by laser pulses. (a) is the electronic dynamic processes of the first plateau including the conduction band C1 and Valence band. (b) and (c) are the electronic dynamic processes of the second plateau including conduction bands C2 or C3 and Valence band, respectively. The starting times of transitions between neighbour conduction bands are labeled by \emph{t} values.}\label{Fig4}
\end{figure}

First, we study the harmonic spectra and field dependence of the cutoff under mid-infrared laser pulses. The electric field strength and the wavelength of the driving laser pulses range from 0.002 a.u. to 0.006 a.u. and 2 $\mu$m to 5 $\mu$m, respectively. The harmonic spectra calculated by Bloch basis agree well with the results by \emph{B}-spline basis in the coordinate space \cite{Guan,Wu}, which is shown in Fig. \ref{Fig2}(a) and (b). One can also find that the cutoffs calculated by the quasi-classical model (marked by the arrows in Fig. \ref{Fig2}(a) and (b)) agree with the results of TDSE well. Experimentally, $\textit{Ghimire et al.}$ had observed the linear dependence of cutoff on the strength of laser fields \cite{Ghimire}. Theoretically, the results show that the cutoff frequency depends linearly on both the electric field strength and wavelength \cite{Schubert,Guan,Wu}. Two or multi-plateau structure in HHG has been predicted theoretically \cite{Wu,Guan}, but not observed in experiments on crystals below the damage threshold \cite{Ghimire,Schubert}. However, in our previous work \cite{Du}, we predicted theoretically that the second plateau may be enhanced by two-three orders in inhomogeneous fields. Multi-plateau structure in HHG from solid rare gases has been confirmed experimentally recently \cite{Ndabashimiye}. It is necessary to extend our proposed quasi-classical model to study higher plateau in HHG by involving multibands.
In Fig. \ref{Fig2}(e) and (f), we show the intuitive insights of the cutoff frequency depending linearly on the electric field strength. The results by the quasi-classical model and TDSE agree well. For each electric field strength, the quasi-particle wave packet has a maximum displacement $\emph{k}(t)_{max}$ in the \emph{k} space. For the first plateau, the cutoff energy is determined by the band gap between the conduction band C1 and valence band at the $\emph{k}(t)_{max}.$ Since the band gap has a linear dependence on \emph{k} values shown in Fig. \ref{Fig2}(c), the orders of cutoffs depend linearly on the $E_{0}$ shown in Fig. \ref{Fig2}(e). Analogously, the band gap between the conduction band C3 and valence band is linearly dependent on \emph{k} approximately shown in Fig. \ref{Fig2}(d). As a result, the cutoff frequency of the second plateau has a linear dependence on the amplitude of laser fields shown in Fig. \ref{Fig2}(f). We also show the wavelength dependence of the cutoffs in Fig. \ref{Fig2}(g) and (h). One can clearly observe the linear wavelength dependence of the cutoffs for the same reason. However, the linear dependence of cutoff on $A_{0}$ is determined by the details of band structure. If the band gap between the conduction bands and valance band is not linearly dependent on \emph{k} in other solid materials, the linear cutoff law in HHG will be broken.

In order to further verify this above-mentioned quasi-classical model, the underlying information in TDSE simulations must be extracted. In Fig. \ref{Fig3}, we perform a time-frequency analysis \cite{Chandre} of HHG calculated by the above introduced TDSE solutions. In Fig. \ref{Fig3}(b) and (c), the color maps represent the time-frequency analysis of the second and the first plateaus, respectively. The curves $\eta(t)$ are the predictions of HHG emission time by the quasi-classical model in Eq. (\ref{E2}). One can find that the emission times of HHG predicted by the quasi-classical model and the results in TDSE agree well. In the subcycle timescale, one can find the harmonic emission law. HHG photons are emitted instantaneously with the motion of quasi-particle wave packet and twice independently in a half cycle. It shows that the amplitude and waveform of laser fields control sensitively the wave packet dynamics and the harmonics emission processes. According to the emission times of the two plateaus in Fig. \ref{Fig3}(b) and (c), one can find the second plateau has a time delay compared with the first plateau. The time delay indicates that the populations of the higher conduction bands are pumped from the nearest conduction band by the driving laser fields. In Fig. \ref{Fig3}(d), the time delays of population on conduction bands agree with that in the time-frequency analysis well.
The details of wave packet dynamics are illustrated in Fig. \ref{Fig4}. In Fig. \ref{Fig4}(a), it presents the wave packet dynamics of the first plateau within three steps: tunneling, Bloch electron oscillation in conduction band C1 and electron-hole pair recombination. The Bloch electron has periodic oscillations driven by laser fields and the HHG emissions follow the oscillations. In Fig. \ref{Fig3}(c) and (a), one can find the Bloch electron oscillation is concurrent with harmonic emission in the time range from -1 o.c. to 2 o.c. It suggests that the occurrence of interband transition needs a critical strength of electric field for Zener tunneling.

In our proposed model, the population of higher conduction bands are pumped from the lower band step by step. For the second plateau, one can find the harmonic emission and wave-packet dynamic processes starting from the near center of laser pulses (-0.25 o.c.) from TDSE in Fig. \ref{Fig3}(b) and predicted from the quasi-classical model in Fig. \ref{Fig4}(b). It can be interpreted that the Bloch electron wave packet in the conduction band C1 is driven toward the Brillouin zone boundary (BZB) at the moment -0.25 o.c., shown in Fig. \ref{Fig3}(a), when the electric field strength is near zero and only a small portion of electrons can tunnel to the conduction band C2 in Fig. \ref{Fig4}(b). This may explain the intensity of the second plateau is 4-5 orders lower than that of the first plateau. However, the moments of population on higher bands can be tuned by chirp or changing the carrier-evelope phase (CEP) in few-cycle laser fields to increase the transition rates \cite{Guan} and HHG intensity of the second plateau consequently. Furthermore, the moments of transition between conduction bands are regulated also by the change of the intensity or wavelength of the driving fields. It can also lead to the change of intensity in the second plateau, as shown in Fig. \ref{Fig2}(b). Our proposed model provides a deep understanding why there is a jump of cutoff energy when the laser intensity reaches a critical value as measured in Ref. \cite{Ndabashimiye}.

From the moment at -0.25 o.c., in Fig. \ref{Fig3}(a), electron wave packet in band C2 is driven back to the Brillouin zone center $\Gamma$ within a quarter optical cycle, and the electric field strength has a maximum at the moment of the center of driving fields. So the Bloch electron can be almost fully pumped to conduction band C3 at $\Gamma$, shown in Fig. \ref{Fig4}(c). As a result, the HHG from C2 and C3 merges together, without clear third plateau.

One can draw a conclusion that the intensity difference of the two plateaus is determined by the magnitudes of populations on C1 C2, and C3. The CEP or chirp of laser pulses can control electron dynamic processes and change the populations of high-lying conduction bands, so as to regulate the intensity of the second plateau. It can be used to interpret the experimental and theoretical results in Refs. \cite{Luu,Guan}.

In summary, this work introduces a quasi-classical three-step model in the \emph{k} space to understand intuitively the Bloch electron dynamics in HHG. It can be used to interpret the phenomenon that HHG cutoffs depend linearly on both electric filed strength and wavelength of the laser pulses. It also explains why the intensity of the second plateau is very weak compared to the first plateau. It provides a promising way to realizes structural reconstruction of energy band by measuring the cutoffs. This model can deliver a prediction of harmonic emission for given solid-state materials and laser fields. This model has been confirmed by the TDSE calculations in the \emph{k} space and coordinate space. This model can also predict the emission time of HHG.  The details of dynamic processes in a subcycle timescale provide a powerful scheme to control HHG trajectories from the solids.

The authors thank Mu-Zi Li and Xin-Qiang Wang very much for helpful discussions. This work is supported by the National Natural Science Foundation of China (Grants No. 11404376 and No. 11561121002).

\end{document}